\documentclass[12pt]{article}
\usepackage{graphicx}

\begin{document}

Preprint \hfill   \hbox{\bf SB/F/01-293} \hrule \vskip 2cm
\pagestyle{empty}

\centerline{ {\large \bf Rindler Particles and Classical
Radiation}} \vskip 1cm \centerline{D.E.D\'{\i}az$^{1}$ and
J.Stephany$^{1,2}$} \vskip .5cm \hskip 3.5cm {\small \it  $^1$
Departamento de F\'{\i}sica, Universidad Sim\'on Bol\'{\i}var,}

\hskip 3.7cm {\small \it Apartado Postal 89000, Caracas 1080-A,Venezuela}
\vskip .2cm
\hskip 3.5cm {\small \it $^2$  Centro de F\'{\i}sica, Instituto Venezolano
Investigaciones Cient\'{\i}ficas,}

\hskip 3.7cm {\small \it Apartado Postal 21827, Caracas 1020-A,
Venezuela} \vskip .2cm \hskip 3.7cm {\small \it
{ddiaz@fis.usb.ve,stephany@usb.ve}}

\vskip 2cm
\centerline{\bf Abstract}
\vskip 0.5cm
 We describe the
 quantum and classical radiation by a uniformly
 accelerating point source in terms of the
 elementary processes of absorption and emission of  Rindler
 scalar photons of the Fulling-Davies-Unruh bath observed by a
 co-accelerating observer.To this end we compute the emission rate
 by a DeWitt detector of a Minkowski scalar particle
 with defined transverse momentum per unit of proper time of the source
 and we show that it corresponds to  the induced absorption or spontaneous
 and induced emission  of Rindler particles from the thermal bath. We then
 take what could be called the inert limit of the DeWitt detector by
 considering the limit of zero gap energy. As suggested by DeWitt, we
 identify in this limit  the detector with a classical point source and
 verify the consistency of our computation with the classical result.
 Finally, we study the behavior of the emission rate in D space-time
 dimensions in connection with the so called apparent statistics inversion.

\vskip 1cm

 \vskip 4cm \hrule
\bigskip
\centerline{\bf UNIVERSIDAD SIMON BOLIVAR} \vskip .5cm \vfill

\section{Introduction}
\hspace{3mm} The discussion of the radiation from a uniformly
accelerated charge as observed from different reference frames  is
the last surprise in Peierls's book \cite{Pei79}. Even from a
classical point of view this topic has been the subject of many
controversies in the past and a satisfactory solution requires a
careful analysis of the space-time radiation distribution
\cite{Pei79,Bou80}. In this note we are interested in relating
some aspects of the radiation process described from a
semi-classical point of view with the classical radiation pattern.
In the semi-classical approach the quantized field is   coupled
linearly to a classical source. The radiation process results from
the emission of what is convenient to call Minkowski particles, at
a rate computed according to the standard rules of Quantum Field
Theory. Alternatively, there is the  inequivalent quantization
scheme \cite{Ful73,Bir82} using Rindler \cite{Rin77} coordinates
which gives the natural description for a co-accelerating Rindler
observer.  For a Rindler observer the forward and backward light
cone surfaces appear as horizons. For this reason this scheme is
restricted to the outer region known as the Rindler wedge. The
fundamental excitations  or Rindler particles are not equivalent
to the usual Minkowski particles. In particular the Minkowski
vacuum is a mixture of Rindler particles with a thermal
distribution which is one of the way to look at the Unruh effect,
the other being the study of the excitation of an accelerated
detector in Minkowski vacuum.

It is natural then to ask for a direct interpretation of the
radiation process in terms of Rindler particles. This has been
done in several works \cite{Wal84,Gin87,Hig92,Higu92,Ren94}. Here
we are particulary interested in the approach of
Ref.\cite{Hig92,Higu92,Ren94} where it was shown that the
classical rate of particle emission computed in the inertial frame
can be reproduced in the co-accelerating frame where, taking into
account the Fulling-Davies-Unruh thermal bath it is shown to be
equivalent to the combined rate of emission and absorption of
zero-energy Rindler particles. This result was calculated in the
charge rest frame by Higuchi, Matsas and Sudarsky
\cite{Hig92,Higu92}, for the electromagnetic field, and by Ren and
Weinberg \cite{Ren94}, for the scalar field by considering in a
perturbative scheme a classical pointlike source following an
hyperbolic trajectory coupled to the field. In both cases the
explicit computation requires applying a regularization procedure
a fact that can be understood because the static source in Rindler
coordinates  can only excite zero-frequency Rindler particles and
then there should be no induced emission but in compensation the
density of quanta in the Fulling-Davies-Unruh bath becomes
infinite as the frequency goes to zero and therefore, a regulator
is needed to carry out the intermediate calculation. In
\cite{Hig92,Higu92,Ren94} the regulator is taken to be the
frequency of a fictitious oscillation of the charge amplitude
which at the end of the calculation is made go to zero.

In the present note we show that this result can be related to the
computation of the emission rate in the presence of an
Unruh-DeWitt detector. We express the emission rate for any value
of the energy shift $E$ of the detector and we show that for $E$
going to zero we recover de classical result. We begin discussing
the radiative process for the Unruh-DeWitt detector in the
inertial reference frame. Then, we compute this same emission rate
of a Minkowski scalar photon from the point of view of the
co-accelerating observer using the mode expansion of the Green
function in terms of the Rindler particles. By this way we can
directly express the emission rate of a Minkowski scalar photon as
a combination of the Rindler particle emission and absorption
rates weighted by thermal factors. We then make the connection
with the radiation pattern of a classical source and with the
results of \cite{Hig92,Higu92,Ren94} by taking a limiting
procedure, that we shall call the inert limit. It consists in
suppressing  the internal structure of the detector by letting the
energy of the excited state $E$ go to zero. This follows an
observation made by DeWitt~\cite{DeW79} and used later by
Kolbenstvedt~\cite{Kol88} in a purely inertial description of the
excitation of the DeWitt detector.

\section{Radiative processes for the DeWitt monopole detector}

\hspace{3mm} Consider a DeWitt monopole detector \cite{DeW79} with
uniform acceleration $a$, that is a point-like object with
internal energy levels, describing the trajectory
\begin{equation}
t(\tau)=\frac{1}{a}\sinh{a\tau},\quad x(\tau)=\frac{1}{a}\cosh{a\tau},
\quad y(\tau)=z(\tau)=0,
\end{equation}
where $\tau$ is the proper time. Suppose that the  detector is
linearly coupled to the quantized real massless Klein-Gordon field
via the interaction Lagrangian given by
\begin{equation}
L_{int}=m(\tau)\phi[x^{\mu}(\tau)].
\end{equation}
Here $x^{\mu}(\tau)$ is the world line of the detector and
$m(\tau)$, is the monopole operator associated to the detector.
The free evolution of a monopole matrix element is given by
\begin{equation}
\langle \mathcal{E}|m(\tau)|\mathcal{E}'\rangle =\langle
\mathcal{E}|m(0)|\mathcal{E}'\rangle e^{i(\mathcal{E-E'})\tau}.
\end{equation}

Now, let us write the probability of excitation of the detector
from the ground state to a final excited state of energy
$\mathcal{E}$, starting with the field in the Minkowski vacuum and
summing over all the final states of the field.  To lowest order
in perturbation theory it is given by
\begin{equation}
\mathcal{P}(E)=\left|\langle \mathcal{E}|m(0)|0\rangle
\right|^2\;\mathcal{F}(E),
\end{equation}
where the response function $\mathcal{F}(\mathcal{E})$ is the
Fourier transform of the positive-frequency Wightman function of
the field, with respect to the proper time~\cite{DeW79,Bir82}
\begin{equation}
\mathcal{F}(E)\equiv\int^\infty_{-\infty}d\tau
\int^\infty_{-\infty}d\tau'
e^{iE(\tau-\tau')}D^+[x(\tau),x(\tau')].
\end{equation}
The other factor is called the selectivity of the detector and
depends only on its internal structure. For convenience we have
set $E$ as the energy difference between the final state of the
detector and its initial state, by letting it be negative we have
the de-excitation of the detector and both cases can be treated on
equal footing with $E=\pm \mathcal{E}$.

For the uniform accelerating trajectory the
Wightman function is stationary, i.e., it depends only on the
proper time difference $\sigma\equiv\tau-\tau'$ and we write
$D^+[x(\tau),x(\tau')]=D^+[\sigma]$. So  the
response function per unit proper time interval can be
computed by factoring out
$T\equiv\int^\infty_{-\infty}d\overline{\tau}$, where
$\overline{\tau}\equiv\frac {\tau+\tau'}{2}$.
\begin{equation}
\frac{\mathcal{F}(E)}{T} =\int^\infty_{-\infty}d\sigma\,
e^{iE\sigma}\,D^+[\sigma].
\end{equation}
Using the standard plane wave expansion of the field, the Wightman
function is given by
\begin{equation}
D^+[x,x']=\langle
0_M|\phi(x)\phi(x')|0_M\rangle=\int\frac{d^3\mathbf{p}}
{16\pi^3\omega_p}e^{-ip(x-x')},
\end{equation}
where $x^{\mu}\equiv(t,\mathbf{x})$,
$p^{\mu}\equiv(\omega_p,\mathbf{p})$ and $\omega_p=|\mathbf{p}|$.
Evaluating on  the hyperbolic trajectory we get,
\begin{equation}
\Delta t=\frac{1}{a} \{ \sinh{a\tau}-\sinh{a\tau'}\}=
\frac{2}{a}\sinh{\frac{a\sigma}{2}\cosh{a\overline{\tau}}},
\end{equation}
\begin{equation}
\Delta x=\frac{1}{a} \{ \cosh{a\tau}-\cosh{a\tau'}\}=
\frac{2}{a}\sinh{\frac{a\sigma}{2}\sinh{a\overline{\tau}}},
\end{equation}
\begin{equation}
\Delta y=\Delta z=0.
\end{equation}
Setting $p^{\mu}=(\omega_p,\kappa,\mathbf{k})$, we
obtain
\begin{equation}
\frac{\mathcal{F}(E)}{T}
=\int\frac{d^3\mathbf{p}}{16\pi^3\omega_p}
\int^\infty_{-\infty}d\sigma\, e^{iE\sigma}\,
e^{\frac{-2i}{a}\sinh{\frac{a\sigma}{2}}\,
\{\omega_p\cosh{a\overline{\tau}}-\kappa\sinh{a\overline{\tau}}\}}.
\end{equation}
After changing variables
$\cosh{\eta}\equiv\frac{w}{k}\cosh{a\overline{\tau}}
-\frac{\kappa}{k}\sinh{a\overline{\tau}}$ in the
integral we have

\begin{equation}
\frac{\mathcal{F}(E)}{T} =\int
d^2\mathbf{k}\int^{\infty}_{-\infty}d\eta
\int^\infty_{-\infty}\frac{d\sigma}{16\pi^3}\, e^{iE\sigma}\,
e^{\frac{-2ik}{a}\sinh{\frac{a\sigma}{2}}\, \cosh{\eta}}.
\end{equation}
At the lowest order, we can read from here the probability of
emission per unit proper time of a Minkowski particle with defined
transverse momentum
\begin{equation}
\label{Minko} \frac{d\,\mathcal{P}(E)}{T
d^2\mathbf{k}}\Big|_{Mink,Emiss} =\left|\langle
\mathcal{E}|m(0)|0\rangle \right|^2\;\int^{\infty}_{-\infty}d\eta
\int^\infty_{-\infty}\frac{d\sigma}{16\pi^3}\, e^{iE\sigma}\,
e^{\frac{-2ik}{a}\sinh{\frac{a\sigma}{2}}\, \cosh{\eta}}.
\end{equation}
In case the detector is prepared in the excited state and decays
to the ground state, the probability is given by the same
expression with negative E .

This result is compared below with the computation in the
co-accelerating reference frame and in particular it is shown to
lead to the same behavior in the classical limit.

Let us now turn our attention to the Rindler frame.  In Rindler
coordinates $(\zeta,\xi,\mathbf{y})$, where $\mathbf{y}$ are the
transverse coordinates,
\begin{equation}
x=\xi\cosh{\zeta}\qquad\qquad t=\xi\sinh{\zeta}
\end{equation}
and
\begin{equation}
ds^2 = -\xi^2d\zeta^2+d\xi^2+d\mathbf{y}^2,
\end{equation}
the hyperbolic motion is given now by $\xi=1/a=const$ and
$\mathbf{y}=\mathbf{0}$, with proper time $\tau=\zeta/a$, (fig.1).
\begin{figure}
\begin{center}
\resizebox{80mm}{!}{\includegraphics{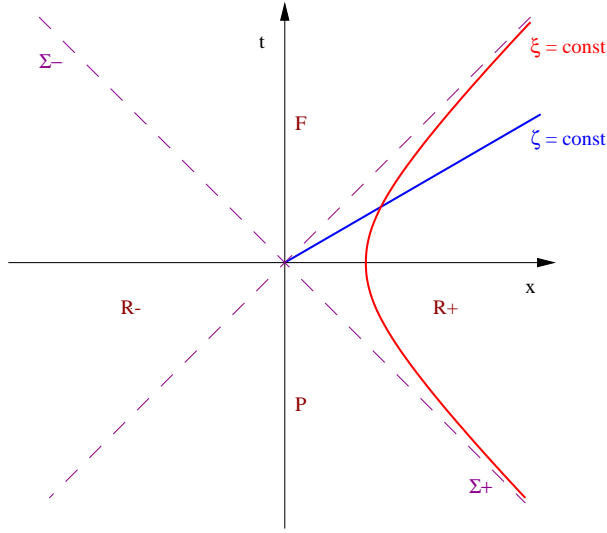}}
\caption{Rindler partition of Minkowski spacetime. The hyperbolic
trajectory is given by $\xi=constant$.}
\label{fig1}
\end{center}
\end{figure}

We can use the Rindler-Fulling expansion of the
field based on the solutions of the Klein-Gordon equation in the
Rindler wedge~\cite{Bir82,Sci81}

\begin{equation}
u_{\nu\mathbf{k}}(x)=\frac{\sqrt{\sinh{\pi\nu}}}{2\pi^2}
e^{-i\nu\zeta+i\mathbf{k}\mathbf{y}}K_{i\nu}(k\xi)\equiv
u_{\nu\mathbf{k}}(\xi,\mathbf{y})e^{-i\nu\zeta}
\end{equation}
and recalculate the Green function for $x,x'$ in the Rindler
wedge. Now we must treat Minkowski vacuum as a many-particle state
of Rindler quanta, therefore the Wightman function is replaced
\cite{Bir82} by
\begin{equation}
D^+[x,x'] =\int_{0}^{\infty}\!d\nu\!\int\!d^{2}\mathbf{k}\!\left\{
 u_{\nu\mathbf{k}}(x)u^*_{\nu\mathbf{k}}(x')\left[1+n_{\nu\mathbf{k}}\right]
 +
 u^*_{\nu\mathbf{k}}(x)u_{\nu\mathbf{k}}(x')\,n_{\nu\mathbf{k}}\right\},
\end{equation}
where the mean number of Rindler quanta in Minkowski vacuum is
given by a thermal distribution at the Unruh-Davies temperature
$T_a=a/2{\pi}$

\begin{equation}
n_{\nu\mathbf{k}}=\frac{1}{e^{2\pi\nu}-1}.
\end{equation}

For the response function on the hyperbolic trajectory,
we then get~\cite{Sci81,Bir82,Gin87}

\begin{eqnarray}
\frac{\mathcal{F}(E)}{T}&=&\int^{\infty}_{-\infty}d\sigma
e^{iE\sigma}\int_{0}^{\infty}d\nu\int d^{2}\mathbf{k} \mid
u_{\nu\mathbf{k}}(1/a,\mathbf{0})\mid^2 \times \nonumber \\
&&\left\{e^{-i\nu a\sigma}\;\left[\frac{1}{e^{2\pi\nu}-1}\right] +
e^{i\nu a\sigma}\;\left[1+ \frac{1}{e^{2\pi\nu}-1}\right]
\right\}.
\end{eqnarray}

After integration we  obtain the following expression for the
probability of emission per unit proper time of a Minkowski
particle with defined transverse momentum to be compared with
(\ref{Minko})

\begin{eqnarray}
\label{phase} \frac{d\,\mathcal{P}(E)}{T
d^2\mathbf{k}}\Big|_{Mink,Emiss} &=& \left|\langle
\mathcal{E}|m(0)|0\rangle \right|^2\; \frac{2\pi}{a}\mid
u_{|E|/a,\mathbf{k}}(1/a,\mathbf{0})\mid^2 \times \nonumber \\
&& \left\{\Theta(E)\;\left[\frac{1}{e^{2\pi E/a}-1}\right] +
\Theta(-E)\;\left[1+ \frac{1}{e^{2\pi |E|/a}-1}\right] \right\}.
\end{eqnarray}

In terms of elementary processes of absorption and emission of
Rindler quanta, this can be written as~\cite{Wal84,Gin87,Sva92}

\begin{eqnarray}
\frac{d\,\mathcal{P}(E)}{T d^2\mathbf{k}}\Big|_{Mink,Emiss} &=&
\Theta(E)\;\left[\frac{1}{e^{2\pi E/a}-1}\right]\frac{d\,
\mathcal{P}(E)}{T d^2\mathbf{k}}\Big|_{Rind,Abs} + \nonumber \\
&& \Theta(-E)\;\left[1+ \frac{1}{e^{2\pi |E|/a}-1}\right]
\frac{d\,\mathcal{P}(E)}{T d^2\mathbf{k}}\Big|_{Rind,Emiss}.
\end{eqnarray}

This gives the relation between the two observers description of
the ``click" of the detector. The emission of a Minkowski particle
with defined transverse momentum with a shift $E$ of the detector
level corresponds to the induced absorption of a Rindler quantum
from the bath when the detector goes up or to the spontaneous plus
induced emission of a Rindler scalar photon to the bath when the
detector goes down (see fig.2 for diagrammatic description of the
situation).

\begin{figure}
\begin{center}
\resizebox{30mm}{!}{\includegraphics{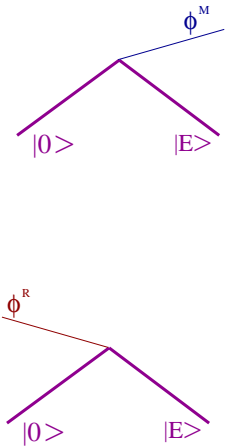}}
\end{center}
\caption{The emission of a Minkowski quantum $\phi^M$ with the
excitation of the detector corresponds to the induced absorption
of a Rindler quantum $\phi^R$ from the FDU thermal bath.}
\label{fig2}
\end{figure}
\noindent The equality between equations (13) and (20) can be
checked out by the explicit use of Nicholson's integral with
imaginary order and argument of the Bessel
functions~\cite{Wat66}).

\section{Classical radiation and the inert limit}

\hspace{3mm} The connection with the radiation of a classical
source is made by following DeWitt's~\cite{DeW79} observation that
if the detector were inert, possessing no internal degrees of
freedom, the pattern of particle emission would be that of a given
accelerating source. Taking  the limit $E \rightarrow 0$ no
structure is left to the detector. In the inertial expression
(\ref{Minko}), the limit is done by setting $E=0$, identifying
$\left|\langle 0|m(0)|0\rangle \right|^2$ with the classical
monopole squared charge  $q^2$ and evaluating the two
integrals~\cite{Gra80}

\begin{equation}
\frac{d\,\mathcal{P}(0)}{T d^2\mathbf{k}}\Big|_{Mink,Emiss}= q^2
\int^{\infty}_{-\infty}d\eta
\int^{\infty}_{-\infty}\frac{d\sigma}{16\pi^3a}
e^{-2i\frac{k}{a}\sinh{\frac{\sigma}{2}}\,\cosh{\eta}} =
\frac{q^2}{4\pi^3 a}\,K^2_0(k/a).
\end{equation}

For the non-inertial computation there are
two competing contributions, the probability of
absorption/emission of a Rindler quantum that  goes to zero
and the termal factor that goes to infinity. Nevertheless the limit is
well defined and gives the same finite result as above. That is for $E
\rightarrow 0^+$

\begin{equation}
\left|\langle E|m(0)|0\rangle \right|^2\;
\frac{2\pi}{a}\frac{\sinh{\pi E/a}}{4\pi^4}\mid K_{iE/a}(k\xi)
\mid^2\;\frac{1}{e^{2\pi E/a}-1}\quad\longrightarrow\quad
\frac{q^2}{4\pi^3 a}\,K^2_0(k/a).
\end{equation}

Here we also make connection with the result of
Ref.\cite{Hig92,Higu92,Ren94} with the advantage of getting the
correct numerical factor (In \cite{Hig92,Higu92,Ren94} an
aditional $\sqrt{2}$ has to be put by hand..

The same result is obtained for $E \rightarrow 0^-$ and therefore
our computation regardless of the initial state of the detector.
We stress that the equivalence in the inert limit is preserved due
to the special role played by the zero-energy Rindler scalar
photons of the bath which are the ones involved in this limit.

The previous calculations can be generalized to more than two
transverse dimensions.  For D-2 transverse dimensions  the response
function in the inert limit gives

\begin{equation}
\frac{F(0)}{T}=\frac{(2\pi^2)^{1-D}}{\Gamma(\frac{D-1}{2})}
\Gamma(\frac{D}{2}-1)^2 .
\end{equation}

This remains finite, regardless the parity of the number of
space-time dimensions ($D>2$). One concludes that the underlying
statistics is of the Bose-Einstein type because a growing number
of quanta in the $E\rightarrow 0$ states is needed in order to
have a finite induced effect. This confirms the arguments of
Ref.~\cite{Unr86,Gin87} in the sense that there is no inversion of
the statistics in odd dimensions as proposed  in the literature.

\section{Conclusion}

\hspace{3mm} The  point we want to stress in this work is that for
an accelerated Unruh-DeWitt detector  the calculations based on
emission/absorption of Rindler quanta in the presence of the
Fulling-Davis-Unruh bath are shown to be equivalent to the
standard ones involving Minkowski quanta. We emphasize that the
behavior of the Unruh-DeWitt detector in the limit of zero energy
gap corresponds to that of a classical source linearly coupled to
the quantum field. In this form we make contact with a previous
approach \cite{Wal84,Gin87,Hig92,Higu92,Ren94} where the classical
radiation pattern was obtained in semiclassical formalism through
and regularization procedure.

In addition, the finite response per unit proper time in the {\em
inert limit} in any $D>2$ spacetime dimensions confirms that the
underlying statistics is of the Bose-Einstein kind, because the
overpopulation of the zero-energy states is necessary in order to
wind up with a finite induced emission probability.

\subsection*{Acknowledgment}
We thank V\'{\i}ctor Villalba and Nami Fux Svaiter for useful
discussions.

\end{document}